\documentstyle[aps,preprint]{revtex}
\begin{document}
\hsize = 7.0in
\widetext
\draft
\tighten

%\begin{document}

\preprint{EFUAZ FT-94-10}

\title{A Note on the Majorana Theory for $j=1/2$ and $j=1$
Particle States\thanks{Submitted to  the XVIII
Oaxtepec Symp. on Nuclear Physics. 4-7 de enero de 1995.}}

\author{Valeri V. Dvoeglazov\thanks{On leave of absence from
{\it Dept. Theor. \& Nucl. Phys., Saratov State University,
Astrakhanskaya ul., 83, Saratov\, RUSSIA.} Internet
address: dvoeglazov@main1.jinr.dubna.su}}

\address {
Escuela de F\'{\i}sica, Universidad Aut\'onoma de Zacatecas \\
Antonio Doval\'{\i} Jaime\, s/n, Zacatecas 98000, ZAC., M\'exico\\
Internet address:  VALERI@BUFA.REDUAZ.MX}

\date{December 30, 1994}

\maketitle

\begin{abstract}

The wave equations for self/anti-self conjugate
Majorana-McLennan-Case $j=1/2$ and $j=1$ spinors,
proposed by Ahluwalia, are re-written to covariant form.
The connection with the Foldy-Nigam-Bargmann-Wightman-Wigner
(FNBWW) type quantum field theory is discussed.
\end{abstract}


\newpage

Thanks to electroweak theory
we have accustomed to thinking
of a neutrino $\nu$ and its antineutrino $\bar \nu$
as distinct particles. A number of precise experiments confirmed the
Weinberg-Salam-Glashow model. However, the present situation
in neutrino physics seems to me to be not satisfactory.
Modern experiments and observations brought
many ``black spots" at the cloudless sky of the Standard Model (SM).
They are: the solar neutrino puzzle,
the negative mass squared problem, the atmospheric neutrino anomaly,
speculations on the neutrinoless double $\beta$- decay,
the tentative experimental evidence for a tensor coupling
in the $\pi^- \rightarrow e^- +\bar \nu_e +\gamma$ decay,
the dark matter problem, gamma-ray bursts etc.
Therefore, searches of the models
beyond the framework of the SM  have some reasons.

As long as 1937 Majorana~\cite{Majorana},
Racah and Furry realised that it is possible
to build symmetrical theory in which there is no
difference between neutrino and its antineutrino.
The essential ingredient of that theory is  projecting
the Dirac wave function into self/anti-self conjugate states
that are supposed to answer for truly neutral particles.
The interest in such the models is revived periodically
in connection with important experimental observations
(e.~g, the discovery of the parity violation in weak processes in
the end of the fifties) or with attractive theoretical constructs
(e.~g., Majorana models of neutrino follow from grand unification
theories). The important reformulation of the Majorana theory
has been undertaken by McLennan and Case~\cite{MLC}.
The excellent pedagogical review of the topics related with Majorana
(and his successors) ideas has been presented by Mannheim~\cite{Mannheim}.
Models proposed by Sokolov~\cite{Sokolov} and Doi {\it et al.},
ref.~\cite{Doi}, also deserve a certain attention.

Recently a new concept of Majorana-like fields has
been proposed~\cite{DVA}. Namely, this concept is
based on the use of the type-II $(j,0)\oplus (0,j)$
self/anti-self conjugate spinors\footnote{Under type-I spinors
one understands the Dirac bispinors. Through this paper
we use  terminology and notation of the author of ref.~\cite{DVA}.
The details of the formalism can be found in the cited papers.}
\begin{equation}
\lambda (p^\mu) \equiv \pmatrix{\left (\zeta_\lambda \Theta_{[j]}\right )
\phi_L^* (p^\mu)\cr \phi_L (p^\mu)},\quad
\rho (p^\mu) \equiv \pmatrix{\phi_R (p^\mu)\cr
\left (\zeta_\rho \Theta_{[j]}\right )^*\phi_R^* (p^\mu)}\quad.
\end{equation}
Phase factors $\zeta_\lambda$ and $\zeta_\rho$ are
fixed by the conditions of self/anti-self $\theta$- conjugacy:
\begin{equation}
S^c_{[1/2]}\lambda (p^\mu) =\pm \lambda (p^\mu),\quad
S^c_{[1/2]}\rho (p^\mu) =\pm \rho (p^\mu)\quad,
\end{equation}
for a $j=1/2$ case; and
\begin{equation}
\left [\Gamma^5 S^c_{[1]}\right ] \lambda (p^\mu) =
\pm \lambda (p^\mu)\quad,\quad
\left [\Gamma^5 S^c_{[1]}\right ] \rho (p^\mu) =\pm \rho (p^\mu)\quad,
\end{equation}
for a $j=1$ case\footnote{As shown in ref.~[6b]
self/anti-self $j=1$ spinors do not exist in the framework
of the construct proposed by Ahluwalia. Therefore, the operation
$\left [\Gamma^5 S^c_{[1]}\right ]$ has been defined there.}.
The operators of the charge conjugation are defined
as\footnote{Let us note that the definition of
the charge conjugation matrix for
a $j=1$ case corresponds to the FNBWW-type
quantum field construct~\cite{DVA0}.
The fact indicated in the previous footnote
is connected with this definition. In the
Weinberg-Tucker-Hammer construct~\cite{Weinberg,Tucker}
it is possible to define self/anti-self $j=1$ conjugate spinors,
see ref.~[8a, p. B1327] for an alternative definition
of the charge conjugation operator.
The various models in the $(1,0)\oplus (0,1)$
representation space were also discussed in
refs.~\cite{Sankar,DVA00,DVO0,DVO}.}
\begin{eqnarray}
S^c_{[1/2]} &=& e^{i\theta^c_{[1/2]}} \pmatrix{0 & i\Theta_{[1/2]}\cr
-i\Theta_{[1/2]} &0\cr} {\cal K} \equiv {\cal C}_{[1/2]} {\cal K}\quad,\\
S^c_{[1]} &=& e^{i\theta^c_{[1]}}\pmatrix{ 0& \Theta_{[1]}\cr
-\Theta_{[1]} &0\cr} {\cal K} \equiv {\cal C}_{[1]} {\cal K}\quad,
\end{eqnarray}
where ${\cal K}$ is the operation of the complex conjugation; and
\begin{equation}
\left (\Theta_{[j]}\right )_{\sigma,\,\sigma^\prime}
= (-1)^{j+\sigma} \delta_{\sigma^\prime ,\, -\sigma}
\end{equation}
is the Wigner operator ($\Theta_{[j]} {\bf J} \Theta^{-1}_{[j]} =
-{\bf J}^*$, with ${\bf J}$ being the angular momentum operator
used in the definition of the Lorentz boosts).
The wave equation for any spin has been given in ref.~[6b,c]
in the instant-front formulation of quantum field
theory\footnote{${\bbox \varphi}$ are the parameters of the Lorentz boost.
In the case of bradyons they are defined by the formulas
(3) of ref.~\cite{DVA0}.}:
\begin{equation}\label{eq}
\pmatrix{-\openone &
\zeta_\lambda exp \left ({\bf J}\cdot{\bbox \phi}\right )
\Theta_{[j]} \Xi_{[j]} exp \left ({\bf J}\cdot {\bbox \phi}\right )\cr
\zeta_\lambda exp \left (-{\bf J}\cdot{\bbox \varphi}\right )
\Xi_{[j]}^{-1} \Theta_{[j]} exp \left (-{\bf J}\cdot {\bbox \varphi}\right )
& - \openone\cr} \lambda (p^\mu) =0\,\, .
\end{equation}
The particular cases ($j=1/2$ and $j=1$) are also
presented there (Eqs. (31) and (32), respectively).
The analogous equation in the light-front formulation could be found
in ref.~[6a]. The $\lambda^S (p^\mu)$
were shown to be
the positive energy solutions with $E=+ \sqrt{m^2 +{\bf p}^2}$,
and $\lambda^A (p^\mu)$ are the negative energy solutions with
$E=-\sqrt{m^2 +{\bf p}^2}$ for both spin-1/2 and spin-1 cases.
However, to re-write those equations to covariant form
is a difficult task.
For instance, an attempt of
the author of the formalism~\cite{DVA}
to put  the equation in the form $(\lambda^{\mu\nu} p_\mu p_\nu +
m \lambda^\mu p_\mu
-2m^2 \openone) \lambda (p^\mu)= 0$ was in a certain sense
misleading. He noted himself: ``it turns out that
[matrices] $\lambda^{\mu\nu}$ and $\lambda^\mu$ do not transform as
Poincar\'e tensors." The aim of the present paper is to explain in what
way the equations for $\lambda (p^\mu)$ and $\rho (p^\mu)$ spinors
are re-written to  covariant form.

The crucial point of deriving Eq. (\ref{eq}) is the generalized
Ryder-Burgard relation for type-II spinors\footnote{In ref.~\cite{DVA0}
the relation $\phi_R (\overcirc{p}^\mu) =\pm \phi_L (\overcirc{p}^\mu)$
for type-I Dirac-like rest spinors has been named as the Ryder-Burgard
relation. Through this paper I also use this
name, but I understand that
similar formulas could be found in
earlier papers and books. Read, e.~g., the paragraphs surrounding
equations (25,26) of Ch. 5, ref.~\cite{Novozhilov}.}:
\begin{equation}
\left [\phi_L^h (\overcirc{p}^\mu) \right ]^* = \Xi_{[j]} \phi_L^h
(\overcirc{p}^\mu)\quad,
\end{equation}
where
\begin{equation}
\Xi_{[1/2]} =\pmatrix{e^{i\phi} & 0\cr
0 & e^{-i\phi}\cr},\quad \Xi_{[1]} = \pmatrix{e^{i2\phi} &0 &0\cr
0&1&0\cr 0& 0& e^{-i2\phi}\cr}\quad,
\end{equation}
$h$ is the helicity; $\phi$ is the azimuthal angle associated  with ${\bf
p}$.  However, from the analysis of the parametrization of rest
spinors (formulas 22a-23c of ref.~[6c]) one can conclude that
another form of the generalized Ryder-Burgard relation is possible.
Namely, the form connecting 2-spinors of the opposite helicity is:
\begin{equation}\label{rbug12}
\left [\phi_L^h
(\overcirc{p}^\mu)\right ]^* = (-1)^{1/2-h} e^{-i(\theta_1 +\theta_2)}
\Theta_{[1/2]} \phi_L^{-h} (\overcirc{p}^\mu)\quad ,
\end{equation}
for a $j=1/2$ case;
and
\begin{equation}\label{rbug1}
\left [\phi_L^h
(\overcirc{p}^\mu)\right ]^* = (-1)^{1-h} e^{-i\delta}
\Theta_{[1]} \phi_L^{-h} (\overcirc{p}^\mu)\quad ,
\end{equation}
for a $j=1$ case ($\delta=\delta_1 +\delta_3$ for $h=\pm 1$ and
$\delta=2\delta_2$, for $h=0$).

Provided that the overall phase
factors of the rest spinors are chosen to be $\theta_1 +\theta_2=0$
(or $2\pi$) in a spin-1/2 case and
$\delta_1 +\delta_3 = 0 = \delta_2$, in a spin-1 case,
the Ryder-Burgard relation is written
\begin{equation}\label{rbu}
\left [\phi_L^h
(\overcirc{p}^\mu)\right ]^* = (-1)^{j-h} \Theta_{[j]} \phi_L^{-h}
(\overcirc{p}^\mu)\quad .
\end{equation}
This choice is convenient for calculations.
The same relations exist for right-handed spinors
$\phi_R (\overcirc{p}^\mu)$
in both a $j=1/2$ case and a $j=1$ case.
By using (\ref{rbu}) and following to the procedure of deriving
the wave equation developed in ref.~\cite{DVA} one can obtain
for a $j=1/2$ case ($\hat p=\gamma^\mu p_\mu$):
\begin{eqnarray}\label{eqq}
\left [i\hat p - m\gamma_5 \right
] \Psi^S_{+1/2} (p^\mu) &=& 0\quad,\quad
\left [i\hat
p + m\gamma_5 \right ] \Psi^A_{+1/2} (p^\mu) =0
\quad,\\ \label{eqq1}
\left [i\hat p + m\gamma_5 \right ] \Psi^S_{-1/2} (p^\mu) &=& 0\quad,\quad
\left [i\hat p - m\gamma_5 \right ] \Psi^A_{-1/2} (p^\mu) =0\quad.
\end{eqnarray}
Here we defined new spinor functions:
\begin{eqnarray}\label{sf}
\Psi^{S}_{+1/2} (p^\mu) = \pmatrix{i\Theta_{1/2}
\left [\phi_L^{-1/2} (p^\mu)\right ]^*\cr
\phi_L^{+1/2} (p^\mu)\cr}
\quad &\mbox{or}& \quad\Psi^{S}_{+1/2} (p^\mu)=
-i\pmatrix{\phi_R^{+1/2} (p^\mu)\cr
-i\Theta_{1/2} \left [\phi_R^{-1/2} (p^\mu)\right ]^*\cr},\\
\Psi^{S}_{-1/2} (p^\mu) = \pmatrix{i\Theta_{1/2}
\left [\phi_L^{+1/2} (p^\mu)\right ]^*\cr
\phi_L^{-1/2} (p^\mu)\cr}
\quad &\mbox{or}& \quad\Psi^{S}_{-1/2} (p^\mu)
= i\pmatrix{\phi_R^{-1/2}(p^\mu)\cr
-i\Theta_{1/2} \left [\phi_R^{+1/2} (p^\mu)\right ]^*\cr},\\
\Psi^{A}_{+1/2} (p^\mu) = \pmatrix{-i\Theta_{1/2}
\left [\phi_L^{-1/2} (p^\mu)\right ]^*\cr
\phi_L^{+1/2} (p^\mu)\cr}
\quad &\mbox{or}& \quad\Psi^{A}_{+1/2} (p^\mu)
= i\pmatrix{\phi_R^{+1/2} (p^\mu)\cr
i\Theta_{1/2} \left [\phi_R^{-1/2} (p^\mu)\right ]^*\cr},\\
\label{sfl}
\Psi^{A}_{-1/2} (p^\mu) = \pmatrix{-i\Theta_{1/2}
\left [\phi_L^{+1/2} (p^\mu)\right ]^*\cr
\phi_L^{-1/2} (p^\mu)\cr}
\quad &\mbox{or}& \quad\Psi^{A}_{-1/2} (p^\mu)=
-i\pmatrix{\phi_R^{-1/2} (p^\mu)\cr
i\Theta_{1/2} \left [\phi_R^{+1/2} (p^\mu)\right ]^*\cr}.
\end{eqnarray}
As opposed to $\lambda (p^\mu)$ and $\rho (p^\mu)$
these spinor functions
are the eigenfunctions of the helicity operator of
the $(1/2,0)\oplus (0,1/2)$ representation space, but they are not
self/anti-self conjugate spinors.

The equations (\ref{eqq},\ref{eqq1}) can also
be obtained by the procedure
described in footnote \# 1 of ref.~[6c] with type-I
spinors ($\Psi=column (\phi_R (p^\mu)\quad \phi_L (p^\mu))$)
if imply that the Ryder-Burgard relation has the form
\begin{equation}
\phi_R (\overcirc{p}^\mu)=\pm i\phi_L (\overcirc{p}^\mu)\quad.
\end{equation}
The equations (\ref{eqq},\ref{eqq1})
have been discussed in the old literature
(e.~g., ref.~\cite{Sokolik}). Their relevance
to the problem of describing the neutrino has been noted in
the cited paper.
By using the formulas relating $\Psi$, Eq. (\ref{sf}-\ref{sfl}),
with self/anti-self conjugate
spinors it is easy to find corresponding equations for spinors
$\lambda (p^\mu)$ and $\rho (p^\mu)$.
For the case of spin-1/2 field we obtain
\begin{eqnarray}\label{eql}
\hat p \lambda^S_{\uparrow} (p^\mu)+
im \lambda^S_{\downarrow} (p^\mu) &=& 0\quad,\quad
\qquad \hat p \rho^S_{\uparrow} (p^\mu)-
im \rho^S_{\downarrow} (p^\mu) = 0\quad,\\
\hat p \lambda^S_{\downarrow}(p^\mu) -
im \lambda^S_{\uparrow} (p^\mu)&=& 0\quad,\quad
\qquad \hat p \rho^S_{\downarrow} (p^\mu)+
im \rho^S_{\uparrow} (p^\mu)= 0\quad,\\
\hat p \lambda^A_{\uparrow} (p^\mu)-
im \lambda^A_{\downarrow} (p^\mu)&=& 0\quad,\quad
\qquad\hat p \rho^A_{\uparrow} (p^\mu)+
im \rho^A_{\downarrow} (p^\mu)= 0\quad,\\ \label{eqll}
\hat p \lambda^A_{\downarrow} (p^\mu)+
im \lambda^A_{\uparrow} (p^\mu)&=& 0 \quad,\quad
\qquad  \hat p \rho^A_{\downarrow} (p^\mu)-
im \rho^A_{\uparrow} (p^\mu)= 0\quad.
\end{eqnarray}
If imply similarly to~[6c] that
$\lambda^S_{\uparrow\downarrow} (p^\mu)$ (and $\rho^A_{\uparrow\downarrow}
(p^\mu)$) are the positive-energy solutions and
$\lambda^A_{\uparrow\downarrow} (p^\mu)$
(and $\rho^S_{\uparrow\downarrow} (p^\mu)$)
are the negative-energy solutions, the equations (\ref{eql}-\ref{eqll})
in the coordinate space can be written
\begin{eqnarray}\label{cr1}
\partial_\mu \gamma^\mu \lambda_\eta (x) +
\wp_{\uparrow\downarrow} m\lambda_{-\eta} (x)&=&0\quad ,\\
\label{cr2}
\partial_\mu \gamma^\mu \rho_\eta (x) +
\wp_{\uparrow\downarrow} m\rho_{-\eta} (x)&=&0\quad ,
\end{eqnarray}
where $\wp_{\uparrow\downarrow}=\pm 1$ with the sign is ``$+$" if
$\eta=\uparrow$ and the sign is ``$-$" if $\eta=\downarrow$. The indices
$\eta$ ($\uparrow$ or $\downarrow$)
should be referred to the chiral helicity introduced in~[6b,p.10].
This form (Eqs. (\ref{cr1}) and (\ref{cr2}))
is very similar to the Dirac equation, however, the sign at the
mass term can be opposite and
the spinors enter in the equations with
opposite chiral helicities. The Dirac equation with
opposite sign at  the mass term had been considered (in different aspects)
in refs.~\cite{Markov,Beli,Brana}. Eqs. (\ref{cr1},\ref{cr2})
should be compared with the new form of
the Weinberg equation for $j=1$ spinors in a coordinate
representation, ref.~\cite{DVA0}.

One can incorporate the same chiral helicity states in equations
by using the identities (48a,b) of ref.~[6c]
\begin{eqnarray}\label{i1}
\rho^S_{\uparrow} (p^\mu) &=& - i\lambda^A_{\downarrow}(p^\mu)\quad,\quad
\rho^S_{\downarrow} (p^\mu) = + i\lambda^A_{\uparrow}(p^\mu)\quad,\\ \label{i2}
\rho^A_{\uparrow} (p^\mu) &=& + i\lambda^S_{\downarrow}(p^\mu)\quad,\quad
\rho^A_{\downarrow} (p^\mu) = - i\lambda^S_{\uparrow}(p^\mu)\quad.
\end{eqnarray}
Thus, one can come to
\begin{eqnarray}\label{sc1}
\hat p \lambda^S_{\uparrow\downarrow} (p^\mu)
+m \rho^A_{\uparrow\downarrow} (p^\mu)&=&0\quad,\quad
\hat p \lambda^A_{\uparrow\downarrow}(p^\mu)
+m \rho^S_{\uparrow\downarrow} (p^\mu) = 0\quad,\\ \label{sc2}
\hat p \rho^S_{\uparrow\downarrow}(p^\mu) +m
\lambda^A_{\uparrow\downarrow} (p^\mu) &=& 0\quad,\quad
\hat p \rho^A_{\uparrow\downarrow} (p^\mu)
+m \lambda^S_{\uparrow\downarrow}  (p^\mu) = 0 \quad .
\end{eqnarray}

It is also useful to note the connection between the type-II spinors
$\lambda (p^\mu)$ and $\rho (p^\mu)$
and the type-I Dirac spinor $\psi^D (p^\mu)$
and the charge conjugate to it $(\psi^D (p^\mu))^c$:
\begin{eqnarray}\label{con1}
\lambda^S (p^\mu) &=& \frac{1-\gamma_5}{2}
\psi^D (p^\mu)+\frac{1+\gamma_5}{2}(\psi^D (p^\mu))^c\quad,\\
\lambda^A (p^\mu)&=& \frac{1-\gamma_5}{2}
\psi^D (p^\mu)- \frac{1+\gamma_5}{2}(\psi^D(p^\mu))^c\quad,\\
\rho^S (p^\mu)&=& \frac{1+\gamma_5}{2}
\psi^D (p^\mu)+\frac{1-\gamma_5}{2}(\psi^D(p^\mu))^c\quad,\\ \label{conl}
\rho^A (p^\mu)&=& \frac{1+\gamma_5}{2}
\psi^D (p^\mu)- \frac{1-\gamma_5}{2}(\psi^D(p^\mu))^c\quad.
\end{eqnarray}
Then the equations (\ref{sc1},\ref{sc2})
could be re-written in the form with the type-I spinors:
\begin{eqnarray}
\left (\hat p +m \right ) \psi^D_{\pm 1/2} (p^\mu) +
\left (\hat p +m\right ) \gamma_5 (\psi^D_{\pm 1/2} (p^\mu))^c &=& 0\quad,\\
\left (\hat p -m \right ) \gamma_5\psi^D_{\pm 1/2} (p^\mu) -
\left (\hat p -m\right ) (\psi^D_{\pm 1/2} (p^\mu))^c &=& 0\quad,\\
\left (\hat p +m \right ) \psi^D_{\pm 1/2} (p^\mu) -
\left (\hat p +m\right ) \gamma_5 (\psi^D_{\pm 1/2} (p^\mu))^c &=& 0\quad,\\
\left (\hat p -m \right ) \gamma_5\psi^D_{\pm 1/2} (p^\mu) +
\left (\hat p -m\right ) (\psi^D_{\pm 1/2} (p^\mu))^c &=& 0\quad.
\end{eqnarray}
So, we can consider the $(\psi^D_h (p^\mu))^c$
(or $\gamma_5\psi^D_h (p^\mu)$, or their sum) as the positive-energy solutions
of the Dirac equation and $\psi^D_h (p^\mu)$
(or $\gamma_5 (\psi^D_h (p^\mu))^c$, or their sum)
as the negative-energy solutions. The field operator can be defined
\begin{equation}\label{fo}
\Psi = \int \frac{d^3 {\bf p}}{(2\pi)^3} \frac{1}{2p_0}
\sum_{h} \left [(\psi^D_h (p^\mu))^c a_h exp (-ip\cdot x)
+ \psi^D_h (p^\mu) b_h^\dagger exp (ip\cdot x)\right ]\quad .
\end{equation}
The similar formulation has been developed by Nigam and
Foldy~\cite{Nigam}.

Let us note a interesting feature.
We can obtain another interpretation ($\psi^D (p^\mu)$ corresponds
to the positive-energy solutions and $(\psi^D (p^\mu))^c$,
to the negative ones) if choose other overall phase factors
in the definitions of the rest-spinors $\phi_L (\overcirc{p}^\mu)$
and $\phi_R (\overcirc{p}^\mu)$,
formulas (22a-23c) of ref.~[6c].
Namely, the signs at the mass term depend on the form of the generalized
Ryder-Burgard relation; if $\theta_1 +\theta_2 =\pi$ the signs
would be opposite.  One can obtain the generalized equations
(\ref{eql}-\ref{eqll}) for the arbitrary choice  of the phase
factor. For $\lambda^S (p^\mu)$ spinors they are following:
\begin{eqnarray}
\cases{i\hat p \lambda^S_{\uparrow} (p^\mu) - m {\cal T}
\lambda^S_\downarrow  (p^\mu) =0&\cr
i\hat p \lambda^S_{\downarrow} (p^\mu) +
m {\cal T} \lambda^S_\uparrow (p^\mu) = 0}\quad,
\end{eqnarray}
where
\begin{equation}
{\cal T} =\pmatrix{e^{i(\theta_1 +
\theta_2)} &0\cr
0& e^{-i(\theta_1+\theta_2)}\cr}\quad.
\end{equation}
In the case $\theta_1 +\theta_2 =\pm {\pi \over 2}$
we also have the correct physical dispersion,
$p_0^2 - {\bf p}^2 = m^2$ for $\lambda (p^\mu)$ spinors.

Next, one can see from (\ref{cr1},\ref{cr2})
that neither $\lambda^{S,A} (x)$ nor $\rho^{S,A} (x)$
are the eigenfunctions of the Hamiltonian operator (we have different
chiral helicity in the ``Dirac" equations). They are not in mass eigenstates.
However $\psi^D$ and $(\psi^D)^c$ are in  mass and helicity eigenstates.
Taking into account the discussion on the positive- and negative- energy
solutions and the interpretation of ``antiparticle" as the particle moving
backward in time~\cite{FS}\footnote{More precisely: in ref.~\cite{Nigam}
it was shown that even without a resort to a plane-wave expansion,
if  the eigenvector
$\phi$ has the eigenvalue ``$-1$" of the normalized
Hamiltonian $\hat H/\vert E\vert$
in the Hilbert space, then $\phi^c$ has the eigenvalue ``$+1$".
This fact leads to the conclusion that the matrix element
$<\lambda^A (0),\,\downarrow \vert \lambda^S (t), \, \uparrow >$ has  the
non-zero value at the time $t$.}
we are ready to put forward the question: can
the high-energy neutrino described by the field
\begin{equation}
\nu^{ML} \equiv \int \frac{d^3 {\bf p}}{(2\pi)^3} \frac{1}{2p_0}
\sum_{\eta}  \left [\lambda^S_\eta (p^\mu) a_\eta (p^\mu)  exp (-ip\cdot
x) + \lambda^A_\eta (p^\mu) a_\eta^\dagger (p^\mu) exp (ip\cdot x) \right ]
\end{equation}
``oscillate" from the state of one chiral helicity to another
chiral helicity (e.~g., $\lambda^S_{\uparrow} \leftrightarrow
\lambda^A_\downarrow$) with the
``oscillation length" being of the order
of the wavelength of de Broglie? Let us recall that
the idea of oscillations of the left-handed neutrino
into its $CP$ conjugate particle has been proposed by
Pontecorvo~\cite{Pontecorvo} long ago.
It is useful to repeat~\cite{DVA} that
$\lambda_{\downarrow }^{S,A}$
and $\rho_{\uparrow}^{S,A}$ are the only surviving spinors
in the massless limit.

For the case spin-1 the situation differs in some aspects.
If accept another formulation of the Burgard-Ryder
relation (\ref{rbu}) one has\footnote{Again, one can obtain the
opposite signs in the equations if imply
$\delta_1 +\delta_3 =\pi$  for  $\phi_L (\overcirc{p}^\mu)$
and,  correspondingly, for $\phi_R (\overcirc{p}^\mu)$.}
\begin{eqnarray}
\gamma_{\mu\nu} p^\mu p^\nu \lambda^S_{\uparrow} (p^\mu) - m^2
\lambda^S_{\downarrow}  (p^\mu)&=& 0 \quad,\quad
\gamma_{\mu\nu} p^\mu p^\nu \rho^S_{\uparrow}  (p^\mu) - m^2
\rho^S_{\downarrow}  (p^\mu)= 0 \quad,\\
\gamma_{\mu\nu} p^\mu p^\nu \lambda^S_{\downarrow} (p^\mu)- m^2
\lambda^S_{\uparrow}  (p^\mu)&=& 0 \quad,\quad
\gamma_{\mu\nu} p^\mu p^\nu \rho^S_{\downarrow} (p^\mu)- m^2
\rho^S_{\uparrow}  (p^\mu) = 0 \quad,\\
\gamma_{\mu\nu} p^\mu p^\nu \lambda^S_{\rightarrow} (p^\mu) + m^2
\lambda^S_{\rightarrow}  (p^\mu)&=& 0 \quad, \quad
\gamma_{\mu\nu} p^\mu p^\nu \rho^S_{\rightarrow}  (p^\mu) + m^2
\rho^S_{\rightarrow}  (p^\mu)= 0 \quad,\\
\gamma_{\mu\nu} p^\mu p^\nu \lambda^A_{\uparrow} (p^\mu) + m^2
\lambda^A_{\downarrow} (p^\mu) &=& 0 \quad,\quad
\gamma_{\mu\nu} p^\mu p^\nu \rho^A_{\uparrow} (p^\mu) + m^2
\rho^A_{\downarrow} (p^\mu)= 0 \quad,\\
\gamma_{\mu\nu} p^\mu p^\nu \lambda^A_{\downarrow} (p^\mu)+ m^2
\lambda^A_{\uparrow} (p^\mu)&=& 0 \quad,\quad
\gamma_{\mu\nu} p^\mu p^\nu \rho^A_{\downarrow} (p^\mu) + m^2
\rho^A_{\uparrow} (p^\mu) = 0 \quad,\\
\gamma_{\mu\nu} p^\mu p^\nu \lambda^A_{\rightarrow} (p^\mu) - m^2
\lambda^A_{\rightarrow} (p^\mu) &=& 0 \quad,\quad
\gamma_{\mu\nu} p^\mu p^\nu \rho^A_{\rightarrow} (p^\mu) - m^2
\rho^A_{\rightarrow} (p^\mu) = 0 \quad .
\end{eqnarray}

There exist the identities analogous to (\ref{i1},\ref{i2}). For instance,
under the choice
of the phase factors as $\delta_1^R +\delta_3^R = 2\pi$,
$\delta_1^L +\delta_3^L=0$ and $\delta_2^L = \delta_2^R -\pi = 0$ we have
\begin{eqnarray}
\rho^S_{\uparrow} (p^\mu) &=& -\lambda^S_\downarrow (p^\mu)\quad,\quad
\rho^S_\downarrow (p^\mu)= -\lambda^S_\uparrow (p^\mu)
\quad,\quad \rho^S_\rightarrow (p^\mu)= +\lambda^S_\rightarrow (p^\mu)\quad,\\
\rho^A_{\uparrow} (p^\mu)&=& + \lambda^A_\downarrow (p^\mu)\quad,\quad
\rho^A_\downarrow (p^\mu) = +\lambda^S_\uparrow (p^\mu)\quad,\quad
\rho^A_\rightarrow (p^\mu)= -\lambda^A_\rightarrow (p^\mu)\quad .
\end{eqnarray}
Therefore,
\begin{eqnarray}
\gamma_{\mu\nu} p^\mu p^\nu \lambda^S_{\uparrow\downarrow\rightarrow}
(p^\mu) + m^2 \rho^S_{\uparrow\downarrow\rightarrow} (p^\mu)&=&0\quad,\quad
\gamma_{\mu\nu} p^\mu p^\nu \lambda^A_{\uparrow\downarrow\rightarrow}
(p^\mu) + m^2 \rho^A_{\uparrow\downarrow\rightarrow} (p^\mu)=0\quad\\
\gamma_{\mu\nu} p^\mu p^\nu \rho^S_{\uparrow\downarrow\rightarrow} (p^\mu)
+ m^2 \lambda^S_{\uparrow\downarrow\rightarrow} (p^\mu)&=&0\quad,\quad
\gamma_{\mu\nu} p^\mu p^\nu \rho^A_{\uparrow\downarrow\rightarrow} (p^\mu)
+ m^2 \lambda^A_{\uparrow\downarrow\rightarrow} (p^\mu)=0\quad
\end{eqnarray}
Applying the relations between type-II and type-I spinors which looks like
similar to (\ref{con1}-\ref{conl})
except for $\rho^S \leftrightarrow \rho^A$ we obtain
\begin{eqnarray}\label{eqb}
\left (\gamma_{\mu\nu} p^\mu p^\nu +m^2 \right )\psi^D (p^\mu) +
\left (\gamma_{\mu\nu} p^\mu p^\nu + m^2 \right )\gamma_5(\psi^D (p^\mu))^c
&=&0\quad,\quad\\
\left (\gamma_{\mu\nu} p^\mu p^\nu -m^2 \right )\gamma_5\psi^D (p^\mu) -
\left (\gamma_{\mu\nu} p^\mu p^\nu - m^2 \right )(\psi^D (p^\mu))^c
&=&0\quad,\quad\\
\left (\gamma_{\mu\nu} p^\mu p^\nu + m^2 \right )\psi^D (p^\mu) -
\left (\gamma_{\mu\nu} p^\mu p^\nu + m^2 \right )\gamma_5(\psi^D (p^\mu))^c
&=&0\quad,\quad\\ \label{eqbl}
\left (\gamma_{\mu\nu} p^\mu p^\nu - m^2 \right )\gamma_5\psi^D (p^\mu) +
\left (\gamma_{\mu\nu} p^\mu p^\nu - m^2 \right )(\psi^D (p^\mu))^c
&=&0\quad.\quad
\end{eqnarray}
This  tells us that $\psi^D$ (or $\gamma_5 (\psi^D)^c$)
should be considered as the negative-energy solutions
of the modified Weinberg equation~\cite{DVA0}
and $(\psi^D)^c$ (or $\gamma_5 \psi^D$) as the positive-energy ones.
Like the equations (\ref{cr1},\ref{cr2}) one can write
\begin{eqnarray}
\gamma^{\mu\nu} \partial_\mu \partial_\nu \lambda_\eta (x)
+ \wp_{S,A}
\lambda_{-\eta} (x) &=& 0\quad,\\
\gamma^{\mu\nu} \partial_\mu \partial_\nu \rho_\eta (x)
+ \wp_{S,A} \rho_{-\eta} (x) &=& 0\quad,
\end{eqnarray}
where $\wp_{S,A} =\pm 1$, the sign is ``$+$"
for positive-energy solution $\lambda^S (p^\mu)$ (or $\rho^S (p^\mu)$)
and the sign is ``$-$", for negative-energy solutions $\lambda^A (p^\mu)$
(or $\rho^A (p^\mu)$).
This refers to the $\eta =\uparrow$ or $\eta=\downarrow$. As
for  $\eta=\rightarrow$ it is easy to see that
the equations (44) and (47) have the opposite sign at mass term.

The presence of $\wp_{\uparrow\downarrow}$ in a $j=1/2$ case
or $\wp_{S,A}$ in a $j=1$ case hints that we obtained the examples of
the FNBWW-type quantum field theory. In fact,
the analysis of the field operator
(\ref{fo}) in the Fock space~[19,6c] revealed
that fermion and its antifermion can
possess same intrinsic parities. Bosons
described by the Eqs. (\ref{eqb}-\ref{eqbl})
are found following to ref.~\cite{DVA0}
to be able to carry opposite intrinsic parities.
However, let us not forget  about other
constructs~\cite{Weinberg,Tucker,DVO}
in the $(1,0)\oplus (0,1)$ representation space.
Their investigations deserve
further elaboration.

\smallskip

%\newpage
{\bf Acknowledgements.}  I greatly appreciate
many useful advises of Profs. D. V. Ahluwalia,
A.~F. Pashkov and Yu. F. Smirnov.
The questions of Prof. M. Moreno
and Prof. A. Turbiner helped me to realise the necessity
of the study of neutral particles.

I am grateful to Zacatecas University for a professorship.

\end{document}